\documentstyle{mn}

\newif\ifAMStwofonts

\ifoldfss
  \ifCUPmtlplainloaded \else
    \NewTextAlphabet{textbfit} {cmbxti10} {}
    \NewTextAlphabet{textbfss} {cmssbx10} {}
    \NewMathAlphabet{mathbfit} {cmbxti10} {} 
    \NewMathAlphabet{mathbfss} {cmssbx10} {} 
  \fi
  \ifAMStwofonts
    \ifCUPmtlplainloaded \else
      \NewSymbolFont{upmath} {eurm10}
      \NewSymbolFont{AMSa} {msam10}
      \NewMathSymbol{\upi}     {0}{upmath}{19}
      \NewMathSymbol{\umu}     {0}{upmath}{16}
      \NewMathSymbol{\upartial}{0}{upmath}{40}
      \NewMathSymbol{\leqslant}{3}{AMSa}{36}
      \NewMathSymbol{\geqslant}{3}{AMSa}{3E}

    \fi
  \fi
\fi 

\ifnfssone
  \newmathalphabet{\mathit}
  \addtoversion{normal}{\mathit}{cmr}{m}{it}
  \addtoversion{bold}{\mathit}{cmr}{bx}{it}
  \newmathalphabet{\mathbfit} 
  \addtoversion{normal}{\mathbfit}{cmr}{bx}{it}
  \addtoversion{bold}{\mathbfit}{cmr}{bx}{it}
  \newmathalphabet{\mathbfss} 
  \addtoversion{normal}{\mathbfss}{cmss}{bx}{n}
  \addtoversion{bold}{\mathbfss}{cmss}{bx}{n}
  \ifAMStwofonts
    \ifCUPmtlplainloaded \else
      \UseAMStwoboldmath
      \makeatletter
      \new@mathgroup\upmath@group
      \define@mathgroup\mv@normal\upmath@group{eur}{m}{n}
      \define@mathgroup\mv@bold\upmath@group{eur}{b}{n}
      \edef\UPM{\hexnumber\upmath@group}
      \new@mathgroup\amsa@group
      \define@mathgroup\mv@normal\amsa@group{msa}{m}{n}
      \define@mathgroup\mv@bold\amsa@group{msa}{m}{n}
      \edef\AMSa{\hexnumber\amsa@group}
      \makeatother
      \mathchardef\upi="0\UPM19
      \mathchardef\umu="0\UPM16
      \mathchardef\upartial="0\UPM40
      \mathchardef\leqslant="3\AMSa36
      \mathchardef\geqslant="3\AMSa3E
    \fi
  \fi
\fi 

\ifnfsstwo
  \DeclareMathAlphabet{\mathbfit}{OT1}{cmr}{bx}{it}
  \SetMathAlphabet\mathbfit{bold}{OT1}{cmr}{bx}{it}
  \DeclareMathAlphabet{\mathbfss}{OT1}{cmss}{bx}{n}
  \SetMathAlphabet\mathbfss{bold}{OT1}{cmss}{bx}{n}
  \ifAMStwofonts
    \ifCUPmtlplainloaded \else
      \DeclareSymbolFont{UPM}{U}{eur}{m}{n}
      \SetSymbolFont{UPM}{bold}{U}{eur}{b}{n}
      \DeclareSymbolFont{AMSa}{U}{msa}{m}{n}
      \DeclareMathSymbol{\upi}{0}{UPM}{"19}
      \DeclareMathSymbol{\umu}{0}{UPM}{"16}
      \DeclareMathSymbol{\upartial}{0}{UPM}{"40}
      \DeclareMathSymbol{\leqslant}{3}{AMSa}{"36}
      \DeclareMathSymbol{\geqslant}{3}{AMSa}{"3E}
    \fi
  \fi
\fi 

\ifCUPmtlplainloaded \else
  \ifAMStwofonts \else 
    \def\upi{\pi}
    \def\umu{\mu}
    \def\upartial{\partial}
  \fi
\fi

\title{Theoretical stellar models for old galactic clusters}
\author[V. Castellani et al.]
       {V. Castellani,$^1$ S. Degl'Innocenti,$^1$ M. Marconi $^{1,2}$ \\
  $^1$ Dipartimento di Fisica, Universit\'a di Pisa, 56100 Pisa, Italy \\
  $^2$ Osservatorio Astronomico di Capodimonte, 80131 Napoli, Italy
}

\pagerange{\pageref{firstpage}--\pageref{lastpage}}
\pubyear{1998}

\begin{document}

\maketitle

\label{firstpage}

\begin{abstract}
We present new evolutionary stellar models suitable for  
old Population I clusters, discussing
both the consequences of the most recent improvements in the input 
physics and the effect of element diffusion within the stellar structures.
Theoretical cluster isochrones are presented, covering  the range
of ages from 1 to 9 Gyr for the four selected choices on the metallicity 
Z= 0.007, 0.010, 0.015 and 0.020.
Theoretical uncertainties on the efficiency of superadiabatic convection
are discussed in some details. Isochrone fitting to the CM diagrams of 
the two well observed galactic clusters NGC2420 and M67 indicates that
a mixing length parameter $\alpha$= 1.9 appears adequate for reproducing 
the observed color of cool giant stars. The problems in matching theoretical 
preditions to  the observed slope of MS stars are  discussed.
\end{abstract}


\section{Introduction}

Evolutionary stellar models represent a key ingredient for any  
investigation about  the evolution of galaxies. Theoretical predictions
concerning  cluster isochrones represent
the only available clock for assessing the age of stellar clusters
in our Galaxy and beyond, allowing a meaningful approach to the 
star formation history in local group galaxies. 
Moreover, the body of theoretical 
prescriptions concerning the evolutionary status of stars with different
ages and chemical compositions is at the  basis of the present
effort for understanding the radiative properties of galaxies
in terms of stellar populations. According to
such evidence, it appears of obvious relevance to rely on 
evolutionary scenarios as detailed and reliable as possible.

In a recent paper (Cassisi et al. 1998, hereinafter Paper I) 
we have revisited
stellar models for old Population II stars, discussing  the effects 
of both the most recent improvements in the input physics and 
of element diffusion within the stellar structures. In this paper
we will use the same evolutionary scenario to extend theoretical 
predictions to stellar  metallicities and ages
suitable for Population I stars. As in Paper I, we will discuss in some
detail the reliability of evolutionary predictions vis-a-vis 
well known uncertainties of the theoretical scenario, as a useful warning 
against an incorrect use of theoretical predictions.

The results of selected evolutionary computations will be presented  
and discussed 
in the next section. Sections 3 will deal with a comparison with 
observational data for the two clusters NGC2420 and M67, taken as 
representatives of the "best observed" galactic clusters in selected  
classes of age. On this basis we will collect observational suggestions
concerning a suitable choice of the mixing length parameter to be used
in constructing models of cool stars, together with some warnings about the 
use of theoretical results. Section 4 will present and discuss 
cluster isochrones for selected choices about  star metallicity.
General conclusions are presented in the subsequent section.

\section[]{Theoretical models}

For the sake of discussion, let us first neglect element diffusion
to recall some relevant  points concerning 
the evolutionary scenario for Population I stars.
The upper panel of Figure 1  shows selected tracks in the 
logL, logTe diagram, as
evaluated by adopting a  ratio of the mixing length to the
local pressure scale height $\alpha$=1.6,  and under the 
alternative assumptions Z=0.007, 0.010,
0.015 and 0.020. Data in this figure show the remarkable similarity
of tracks at fixed mass but different metallicities. As already known,
an increase in the metal content moves ZAMS models toward redder colors 
and fainter magnitudes, 
whereas the difference in temperature between the ZAMS and the 
Red Giant Branch (RGB)  decreases.


\begin{figure}
 \vspace{100pt}
 \caption{
Upper panel:  Evolutionary tracks with $\alpha$=1.6
for selected stellar masses and for the labeled values of the assumed metallicities.
Lower panel: Evolutionary tracks 
for the same masses as in the upper panel but for Z=0.015 
and with the labeled assumptions about the mixing length parameter $\alpha$.
}
\end{figure}

However, one knows that theory gives no firm constraints about
the most appropriate value of $\alpha$, so that it appears
of some relevance to bring to the light the consequences of a variation
of the mixing length parameter within reasonable
limits. This is shown in the lower panel of the same Figure 1,  which reports 
evolutionary tracks given in the upper panel for Z=0.015 but under the three 
alternative choices  $\alpha$= 1.0, 1.6  or 2.2. As already known, one
finds that theory gives
firm predictions about the temperature of ZAMS stars only for stars with 
masses larger or of the order of 1.5 M$_{\odot}$. Below this limit 
the uncertainty
in the temperature appears roughly of the same order of magnitude
of the difference produced in the upper panel of Figure 1 by the variation 
in metallicity. Moreover, the figure reinforces the evidence that theory
gives only marginal constraints on the temperature (and thus on the color) 
of the Red Giant Branch (RGB), which appears as a free parameter to be
adjusted -in the case- by tuning the assumption on $\alpha$.
 
To allow the comparison with the observed C-M diagrams, one has to use  
evolutionary tracks to predict cluster isochrones in the chosen bands
of magnitudes and colors. This has been done by first producing
"theoretical" isochrones in the logL, logTe diagram, to be finally
transformed into C-M diagrams according to suitable assumptions
about the bolometric corrections and the color temperature relations.
Let us notice that the intermediate step represent 
the "true" evolutionary result, to be compared with the result of 
similar evolutionary evaluations before the intervention of further 
assumptions about model atmosphere computations.  

Figure 2 shows a selected sample of the new theoretical isochrones, as 
computed for the labeled assumptions about the cluster ages, $\alpha$=1.6
and  Z=0.02. 
One may notice the mild variation of the luminosity of the
clump of He burning stars for cluster ages larger than 1 Gyr,
already discussed in Castellani, Chieffi \& Straniero (1992, hereinafter CCS).
As a consequence, for each given chemical composition and in the above 
quoted range of ages the difference in magnitude between the clump
and the MS turn off appears a good indicator of the cluster ages and,
for each age, the clump luminosity appears as a relevant standard candle
for deriving the cluster distance modulus. Note that for ages lower than 1 
Gyr the luminosity of the clump follows the variation of the size
of the He core through and beyond the "Red Giant Transition" as already
exhaustively discussed in the current literature (see, e.g., Sweigart, 
Greggio \& Renzini 1990).

\begin{figure}
 \vspace{100pt}
 \caption{
 Cluster isochrones in the theoretical plane 
for Y=0.27, Z=0.02, $\alpha$=1.6  and for the selected labeled ages.
}
\end{figure}


Isochrones with solar composition 
(Z=0.02) allow a comparison with  previous CCS results, as obtained from the
same code with the same assumption about the mixing length, but with "old"
input physics. As a most significant 
comparison we will chose  a cluster age large enough
for red giants undergoing electron degeneracy. The comparison, as given
in Figure 3 (upper panel) for the 4 Gyr isochrones, shows only
marginal differences. However, the use of more recent  evaluations of colors 
gives larger differences in the predicted CM diagrams.
This is shown in the lower panel of the same Figure 3, which shows
the new 4 Gyr isochrone as translated into the V, B-V diagram 
adopting for MS stars the empirical relations given by Alonso et al. (1996), 
implemented
with Castelli et al. (1997a,b) model atmospheres, shifted by B-V$\simeq$-0.04
to match the empirical
relations at their limit of validity ($\log{g}\simeq{4}$,
$T_e\simeq{6700}$ K). In this way we attempted to match model atmospheres
to current  observational constraints, though at the cost
of some marginal discontinuities in the predicted run of
colors. Comparison with the isochrone presented by CCS, as given in the
same lower panel, reveals -in particular- that the new transformations
predict a steeper MS, with consequences that will be discussed later
on in this paper.

\begin{figure}
 \vspace{100pt}
 \caption{
 Comparison among selected cluster isochrones from the
present paper and from CCS in the theoretical (upper panel) and
in the observational plane (bottom panel).
}
\end{figure}


\begin{figure}
 \vspace{100pt}
 \caption{
Cluster isochrones as computed for the labeled ages and
chemical composition and under the two labeled assumptions about the
mixing length parameter $\alpha$.
}
\end{figure}


Figure 4 discloses the effect on the isochrones of
different assumptions about the mixing length. As expected on the basis
of evolutionary tracks in Figure 1, one finds that cooler stars are
strongly affected by the mixing length.  MS and TO stars are free 
from such a theoretical uncertainty 
only for B-V smaller than B-V$\simeq$0.4. For color redder than this limit
the uncertainty in color grows, reaching approximately $\delta(B-V)\simeq$0.1
mag. in the interval Mv=4.0-6.0 mag., slowly decreasing at even lower 
MS luminosities.

The above evolutionary
evidence is a consequence of the well known occurrence for which above a 
given effective temperature stellar envelopes become free of 
convection and, thus, not affected by the treatment of such a mechanism. 
Note that decreasing the MS mass, the density of the envelope increases
and convection becomes less and less superadiabatic; as a consequence,
the (convective) less massive MS structures are only marginally
affected by assumptions about the mixing length (vandenBerg, Hartwick \&
Dawson 1983). This  occurrence 
explains the decreasing shift produced  by variation in the mixing 
length on the lower portion of the MS in Figure 4.
The different dependence of
RG and MS stars at a given color is again related to the density of the
envelopes: RG have rarefied envelope, thus experiencing strong 
superadiabaticity, whereas  MS stars have much  
denser envelopes,  thus less affected by superadiabatic convection.

As already noticed by Chaboyer (1995) for Population II systems, 
one finds that the mixing length does affect the calibration of the 
TO magnitudes in terms of cluster ages.  According to the discussion
given in Paper I, the effect of mixing length becomes larger for TO
points located at redder colors, i.e., when 
increasing either the age or the star 
metallicity. Figure 4 shows that such an error is rather negligible for
the 3 Gyr isochrone, but it increases to about 1 Gyr for the 6 Gyr
isochrone, growing up for larger ages. Thus the TO magnitude 
cannot supply age evaluations for old metal rich clusters
more precise than this limit, until an improved treatment of convection 
becomes available for stellar evolutionary models.

Concerning He burning stars, the previous 
Figure 2 shows that for ages larger than, about,
1 Gyr, He burning occurs near the RG branch, in  agreement
with observational evidences. However, the beautiful
theoretical constraint given by the rather constant 
luminosity of the clump of 
He burning stars, will  give less firm predictions  when transferred
into the CM diagram,  since magnitudes
depend on the bolometric correction which, in turn, depends on
the color of the  clump and, thus, on the assumptions adopted
for the mixing length parameter.

Before closing the discussion about the mixing length, let us
here advise the reader that the problem is not, or at least it could
not be,  to find out the "right value" of the mixing length.
In  fact, the mixing length theory is only a rough
(though useful) parametrization of the efficiency of convection, and 
there is no reason for constraining the mixing length parameter to 
be the same in stellar structures characterized not only by different
stellar masses or chemical compositions, 
but also in different  evolutionary phases of a given star. Thus the 
above discussion about the effects of the mixing length has
to be taken as an investigation of the range of theoretical
uncertainty on the various evolutionary phases, without
necessarily assuming that, e.g., MS and RG should have
effective temperatures corresponding to a common value of $\alpha$.

Finally, one may investigate how far the
efficiency of element sedimentation affects the above results.
Figure 5 shows selected isochrones for Z=0.015 as compared with similar 
computations but with element diffusion taken into account. As expected,
one finds that diffusion plays a role only at larger ages, decreasing 
the color of the turn-off (TO), with minor influence on other evolutionary 
phases. It turns out that the diffusion sensitively
affects only the isochrones which are already affected by much larger
uncertainties in the mixing length, masking the effect of
diffusion. Thus one can safely use canonical, no-diffusion models,
bearing in mind the above discussed theoretical indeterminations. 
Note that this is not the case for old  metal poor globular
cluster stars, where diffusion has been efficient for much larger times
(see Paper I).

\begin{figure}
 \vspace{100pt}
 \caption{
 Cluster isochrones for the given chemical composition and
mixing length parameter, but with or without allowing for the
efficiency of element diffusion.
}
\end{figure}


\section{ From observations to theory and back}

According to the discussion given in the previous section one should feel
reluctant to present a theoretical scenario based on a given assumption
about the value of the mixing length parameter. In this section we will
follow a different approach, comparing the evolutionary
scenario depicted in the previous section with observational data,
looking for  more  light about reasonable
theoretical predictions. For accomplishing this goal, we 
focused our attention on two
"best  studied" galactic clusters, namely NGC2420 and M67, for 
which good C-M diagrams have been already presented in the literature.
In the following we will investigate the agreement between
observational data and  theoretical predictions, as computed 
by including element diffusion, searching for observational
constraints to the theory.

\subsubsection{NGC 2420}

Figure 6 shows the  beautiful CM diagram of NGC2420 presented
by Antony Twarog et al. (1990).  Following these authors 
the cluster should have a reddening not smaller than E(B-V)=0.05,
with some evaluations as large as E(B-V)= 0.14 (Cohen 1980).
Thus the color of TO stars should be lower than B-V=0.35.
According to the discussion given in the previous section, we
note that the cluster appears young enough to have TO stars
only marginally affected by the mixing length parameter. Such a fortunate
occurrence decreases the degrees of freedom and made the cluster our
first choice for fitting theory to observations.

\begin{figure}
 \vspace{100pt}
 \caption{
The CM diagram presented by Antony Twarog et al. (1990) for the
galactic cluster NGC2420 and the best fit of present isochrones.
}
\end{figure}


According to Antony Twarog et al. (1990) metallicity estimates give
for the cluster [Fe/H]=-0.35 $\pm$0.10, but with evaluations reaching
[Fe/H]=-0.6 (Canterna et al.  1986). We will assume Z=0.007 ([Fe/H=-0.4]).
By taking Y=0.23 for old metal poor stars and Y=0.27 for the Sun
(Z$_{\odot}$=0.02), a linear interpolation on metallicity finally gives for
NGC2420 Z=0.007 and Y=0.244. 

For fitting theory to observations, taking into account
the already discussed uncertainty on the location of cool stars,
we devised a procedure which runs as follows. We first 
determined  the predicted
location in the CM diagram of the  "He clump" for a suitable range of
assumptions about the mixing length parameter, namely $\alpha$= 1.0-2.2 
As an exemple, this is shown in Figure 7 for
a 1.5 Gyr isochrone, with Y=0.27, Z=0.007.
By forcing the match between the predicted and the observed location
of  both TO and He burning stars 
one eventually finds the cluster reddening, the distance modulus,
and the mixing length value suitable for the observed color of
luminous red stars.

\begin{figure}
\vspace{100pt}
\caption{
Comparison of stellar isochrones for the given age and
chemical composition but for the two labeled values of $\alpha$. The dashed
line shows the predicted locus of the He burning clump when varying 
$\alpha$ between 1 and 2.2.
}
\end{figure}


As disclosed by  the same Figure 6, 
the fitting is not only possible, but it appears
remarkably good, since it reproduces to a high degree of accuracy
the shape of H burning stars near and beyond the overall contraction phases.
In this way one would derive E(B-V)$\sim$0.16, DM$\sim$12.4, $\alpha$=1.9  
and a cluster age of about 1.5 Gyr, in reasonable agreement with previous
CCS results, as obtained for a solar composition. 
Comparison with the isochrones for the ages
1.25 and 1.75 Gyr, as given in the same figure, shows that the formal error
in age should not exceed 0.1 Gyr! Of course, reddening  depends on the adopted relation
between temperatures and colors. Adopting only Castelli et al. (1997) 
models, one would derive
a reddening smaller by 0.05, thus E(B-V)$\sim$0.11. Similarly, the
distance modulus depends on the amount of bolometric
corrections which, however, appears much less model dependent than colors do.

One has finally to notice a disagreement between the theoretical and the
observed location of the lower portion of the main sequence, as due to the 
evident inadequacy of either stellar models or color temperature relations. 
The agreement would be improved adopting for these models 
smaller values of the mixing length, which  means to assume
external convection to be less efficient in MS stars than expected adopting 
a common mixing length ($\alpha$=1.9) for both MS and RG stars. 
However, Figure 7 shows that low masses MS stars are less and less affected
by the mixing length. Thus it appears difficult to reconcile theory with
observation only by tuning the mixing length.  As we will discuss
in the case of M67, we suspect that such a disagreement is due to
the inadequacy of adopted color-temperature relations.

Theoretical predictions can be submitted
to a further independent test. As largely discussed in CCS, 
in relatively young clusters the distribution
of stars in the advanced evolutionary phases is a
rather sensitive function of the cluster age. Thus beyond the agreement
between the CM diagram loci, one can test the theoretical distribution
vis-a-vis the observed distribution. This has been done by computing,
with a Monte Carlo technique, the synthetic CM diagram of the cluster, 
with the same number of off main sequence stars as observed. 
Theoretical predictions, as shown in 
Figure 8, middle panel, appear  in reasonable agreement with 
observations, giving further 
support to the predicted cluster age. The lower panel finally compares
the synthetic cluster with the theoretical isochrone, giving light on the
effect of binary stars on the topology of the overall contraction gap we
 will discuss in the final section.

\begin{figure}
 \vspace{100pt}
 \caption{
The observed CM diagram (upper panel) compared with the
synthetic clusters (middle panel) aiming to reproduce NGC2420 taking
into account the contribution of 30\% of binaries. Comparison with the
isochrone at 1.5 Gyr (bottom panel) shows that binaries tend to fill the 
overall contraction gap. Observational data are not dereddened.
 }
\end{figure}


\subsection{M67}

Figure 9 shows the CM diagram of M67  presented by Montgomery, Marschall \&
Janes (1993). According to these authors, the cluster is characterized by
a metallicity [Fe/H]=-0.05. 
Correspondingly we will refer to an evolutionary scenario as computed for 
Z=0.015. Reddening evaluations range in the interval 
E(B-V)=0.03 - 0.10 (Antony Twarog et al. 1990); in particular Cohen (1980)
analyzed absorption lines of sodium of the interstellar gas giving
E(B-V)=0.09. 

\begin{figure}
 \vspace{100pt}
 \caption{
The color magnitude diagram presented by Montgomery et al.
(1993) for the galactic cluster M67 and the best fit with the 3.25 Gyr isochrone.
}
\end{figure}

Inspection of Figure 9 discloses that now  all the cluster stars lie
in a region affected by the uncertainty about the mixing length. 
Moreover, the MS of the cluster is now largely in the range
of colors where we already found a mismatch between theory and observation. 
The same figure 9 shows that the most luminous portion of cluster stars
can be reasonably fitted by theory  for an age of 3.25 Gyr, 
assuming again $\alpha$=1.9, and with the labeled 
values of cluster distance modulus and reddening. However the location
of the theoretical MS appears far from being satisfactory.

In this context, we have already noticed that new computations
appear in satisfactory agreement with previous CCS evaluation. 
Degl'Innocenti \& Marconi (1998) have also shown that the new theoretical 
4 Gyr isochrone already presented 
in the upper panel of Figure 2 appears also in good agreement with 
similar predictions by Bertelli et al. (1994). Thus theory
appears rather solid in predicting the location in the theoretical HR diagram 
of similar stars, with only a minor influence of the updated physics 
inputs.  Curiously enough, one finds that these isochrones were
already used to give a good fitting of the M67 MS not only by CCS but also by  
Carraro et al. (1996). If we add the evidence that CCS were also nicely 
fitting the MS of NGC2420, one is driven to conclude that the fitting 
of these MS is largely a matter of the adopted color-temperature 
relations and that, unfortunately, more recent predictions of model 
atmospheres are moving theory away from observations.

In this respect one has to notice that recent empiric calibrations of 
V-K colors in terms of stellar temperature, as given by stellar 
interferometry, have revealed that the results of recent model atmospheres
given by Gratton et al. (1996) are supported by observation only
at temperatures around 7000 $^o$K, whereas at lower temperature 
theoretical predictions appear in increasing disagreement with the
calibration (Di Benedetto, private communication).  This suggests
to us that the temperature-color relation is perhaps the 
weak link in the connection theory-observation.
According to these evidences, we are inclined to  conclude that 
the transformation of temperatures into colors is probably
the most relevant problem in using evolutionary theory. 
In this context, theoretical fittings 
as given in Figure 9 should probably be regarded only as an exercise,
whereas a reliable investigation  of cluster evolutionary parameters
should probably wait for a much firmer assessment of this problem

Let us finally notice that 
Montgomery et al. (1993) by fitting CCS isochrones to the same 
observational data already noticed the good agreement between theory and
observations in the MS and TO regions. However, the missed fitting of 
the RG colors
induced the same authors to conclude that "the current generation of 
theoretical isochrones cannot be fit to the observed sequence within the 
observational errors". Even tough one cannot disagree with such a statement, 
we wish here to stress once again that the RG colors is largely matter of 
a cosmetic adjustment of theoretical results, as we have done in
this section.

As already discussed, the computations of the synthetic clusters
can give an independent indication at least of the compatibility of the 
theoretical scenario one is dealing with. Figure 10 shows 
that assuming 10\% of binary stars one finds a satisfactory agreement 
of theoretical predictions with observational data.

\begin{figure}
 \vspace{100pt}
 \caption{
The observed CM diagram for M67 (upper panel) compared
with the synthetic cluster for the labeled values of chemical composition and age
aiming to reproduce M67 taking into account
the contribution of 10\% of binaries, without (middle
panel) and with superimposed the corresponding theoretical cluster
isochrone (bottom panel). Observational data are not dereddened. 
 }
\end{figure}


\section{Cluster isochrones}

The discussion given in the previous sections has shown that 
$\alpha$=1.9 appears the adequate (cosmetic)
choice to produce cluster isochrones for old open clusters, all over
the range of metallicity Z=0.007, 0.015, at least. According to such a 
result, Figure 11 shows cluster isochrones computed under the above quoted
assumptions and for four selected metallicities. Data for all
the computed isochrones are available at the anonymous ftp at
astr18pi.difi.unipi.it (/pub/open),
where we give theoretical (logL, log Te) isochrones together with
V, B-V and V-I magnitudes as predicted according to the
already quoted match between Alonso et al. (1996) and
Castelli et al. (1997 a,b) evaluations.
Bearing in mind the caveat concerning theoretical colors, 
let us now discuss some
theoretical predictions of general relevance.

\begin{figure}
 \vspace{100pt}
 \caption{
Cluster isochrones in the observational plane 
for the labeled assumptions about the chemical composition and
for $\alpha$=1.9. Isochrones are computed including diffusion of
helium and heavy elements. Ages as labeled. Theoretical isochrones are
translated in the observational plane by adopting
the empirical relations given by Alonso et al. (1996), 
implemented with Castelli et al. (1997a,b) model atmospheres,
as described in the text.
}
\end{figure}


Figure 12 gives the bottom luminosity of the clump of He burning stars
as a function of the cluster age in the interval 1-6 Gyr and for
the five explored metallicities. For each given metallicity, the
figure discloses the good constancy of this parameter, which should allow
the use of He burning stars as useful standard candles.

\begin{figure}
 \vspace{100pt}
 \caption{ The bottom luminosity of the clump of He burning
stars (from the isochrones of Figure 11)
 as a function of the cluster age and metallicity.
}
\end{figure}


\begin{figure}
 \vspace{100pt}
 \caption{
 The predicted magnitude of the He clump
(from the isochrones of Figure 11)
 as a function of the metallicity Z for the various explored cluster ages.
}
\end{figure}


However, one 
finds that
for ages larger than 2 Gyr, the luminosity of the clump is slowly 
decreasing when the age increases. Since the clump is in the 
meantime becoming 
redder, the bolometric correction increases, increasing the variation in 
magnitudes.  This is shown in Figure 13, where we report the predicted
(bottom) magnitudes of the clump as a function of the metallicity Z
for selected assumptions about the cluster ages. One finds that 
the linear relation:
\begin{equation}
{\rm Mv}= 1.59 + 0.59~ {\rm log Z}
\end{equation}
reproduces the theoretical results  over the whole range of ages within
0.1 mag. 

As another relevant prediction, Figure 14 gives the difference in magnitude
between the clump and the top luminosity of the H burning sequence (a 
parameter already introduced in CCS) as a function of the cluster ages 
for the chosen metallicities. 

\begin{figure}
 \vspace{100pt}
 \caption{ 
The predicted difference in magnitude between the He
clump and the top of the H burning sequence (from the isochrones of Figure 11) 
as a function of the cluster age and for the labeled assumptions about the cluster
metallicity.
}
\end{figure}


\section{Conclusions}

Galactic stellar clusters have  covered 
a central role in the progress of stellar evolutionary theories, 
giving  direct evidences for  
the evolution  of metal rich, Population I stars for a fairly
large variety of cluster ages and metallicities. However,
throughout this paper we have shown that the situation is far
from being completely satisfactory. Theoretical uncertainties
on the efficiency of external convection play a major role in 
shading a disturbing degree of freedom in relevant theoretical
predictions. Indetermination on both the temperature-color relations and in
the actual cluster reddening add further problems into this
scenario.

In this paper we have attempted an empirical calibration of the theory, 
reaching what we regard as a  satisfactory agreement between
theory and observation. It remains the disturbing discordance
of the MS slope at the larger explored B-V value, whose origin
is far from firmly established. Further improvements in color-temperature
relations and in the determination of cluster reddening are possible
and expected. However external convection keep being
the main outstanding problem. In this sense new 
approaches to the theory of convection, as the one presented by
Canuto \& Mazzitelli (1992) or Lydon, Fox \& Sofia (1993), 
should be carefully tested to well
studied galactic clusters and, if necessary, improved in the hope
of reaching a real knowledge of such a fundamental ingredient of
stellar evolutionary theories. 

As a final point we note that that the predicted luminosity of He
burning clumps is directly correlated with the size of the He
cores at the end of central H burning. Both
NGC2420 and M67 are predicted with Red Giants undergoing electron
degeneracy. The good correspondence
between observations and theory found in the previous section 
supports current evaluations of the physical mechanisms
affecting these structures. Note that in this case, the predicted
He core are only marginally affected by the possible efficiency
of convective overshooting, which however should modulate the
shape of the CM diagram in the Turn Off  region.

In this context Demarque, Sarajedini \& Guo (1994)
have recently proposed an alternative approach to constrain
the efficiency of overshooting, as based on the topology
of the CM diagram in the region of the overall contraction
gap. In our feeling that approach, tough ingenious and theoretically 
ywell founded, suffers  of some limitations mainly due to the 
occurrence of binary stars within the gap (see bottom panels in
the previous Figures 8 
and 10), preventing from firm observational constraints on the
suggested CM parameters.  

\section*{Acknowledgments}

It is a pleasure to thank Giuseppe Bono for a critical reading of
the manuscript and for valuable suggestions. One of the authors, M.M., 
acknowledges the grant from C.N.A.A.

\label{lastpage}


\begin{thebibliography}{99}

\bibitem{} Alonso, A., Arribas, S. \& Martinez-Rogers, C. 1996, A\&A, 313, 873
\bibitem{} Anthony-Twarog, B. J., Twarog, B. A., Kaluzny, J. \& Shara, M. M. 1990,
AJ, 99, 1504
\bibitem{} Bertelli G., Bressan A., Chiosi C., Fagotto F., Nasi E. 1994, A\&AS 106, 275
\bibitem{} Brocato, E., Castellani, V. \& Romaniello M. 1998, in preparation
\bibitem{} Canterna, R., Geisler, D., Harris, H. C., Olszewski, E. \& Shommer, R.
1986, AJ, 92, 79
\bibitem{} Canuto V. \& Mazzitelli I. 1992, ApJ 389, 724
\bibitem{} Caputo F. \& DeSantis,R. 1992, AJ, 104,253
\bibitem{} Carraro G., Girardi L., Bressan A. Chiosi C. 1996, A\&A 305, 849
\bibitem{} Cassisi, S., Castellani, V., Degl'Innocenti, S., Weiss, A., 1998, A\&AS,
in press
\bibitem{} Castellani, V., Chieffi, S., \& Straniero, O. 1992, ApJS, 78, 517
\bibitem{} Castelli, F., Gratton, R. G. \& Kurucz R. L. 1997a, A\&A, 318, 841
\bibitem{} Castelli, F., Gratton, R. G. \& Kurucz R. L. 1997b, A\&A, 324, 432
\bibitem{} Chaboyer B. 1995, ApJ, 444, L9
\bibitem{} Cohen, J. G. 1980, ApJ, 241, 981
\bibitem{} Demarque P., Sarajedini A. \& Guo X-J. 1994, ApJ 426, 165
\bibitem{} Gratton R.G., Carretta E., Castelli F. 1996, A\&A 314, 191
\bibitem{} Lydon T.J., Fox P.A., Sofia S. 1993, ApJ 413, 390
\bibitem{} Montgomery, K. A., Marschall, L. A. \& Janes, K. A. 1993, AJ, 106, 181
\bibitem{} Sweigart A., Greggio L. \& Renzini A. 1990, ApJ 364, 527
\bibitem{} vandenBerg D.A., Hartwick F.D.A. \& Dawson P. 1983, ApJ 266, 747

\end{thebibliography}
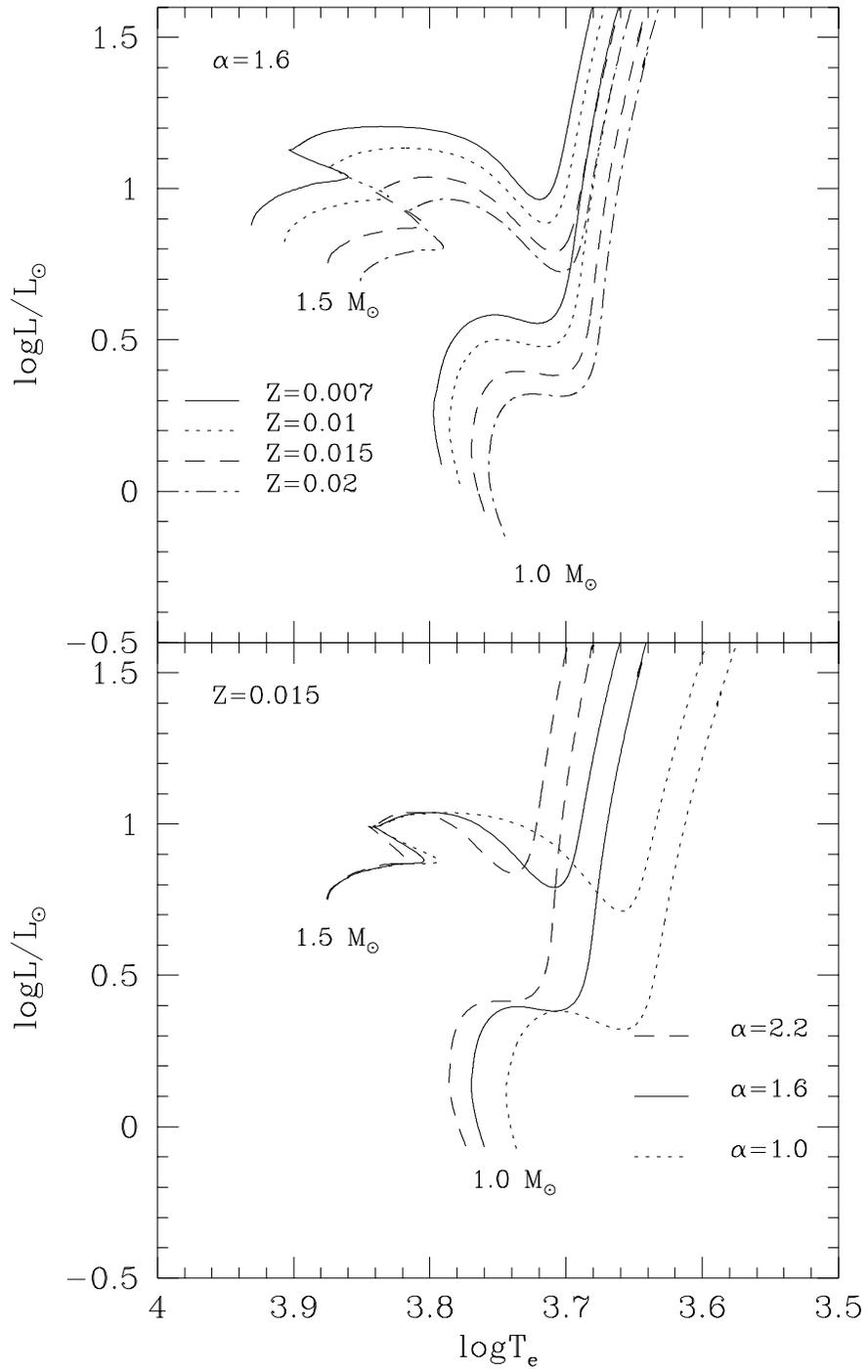
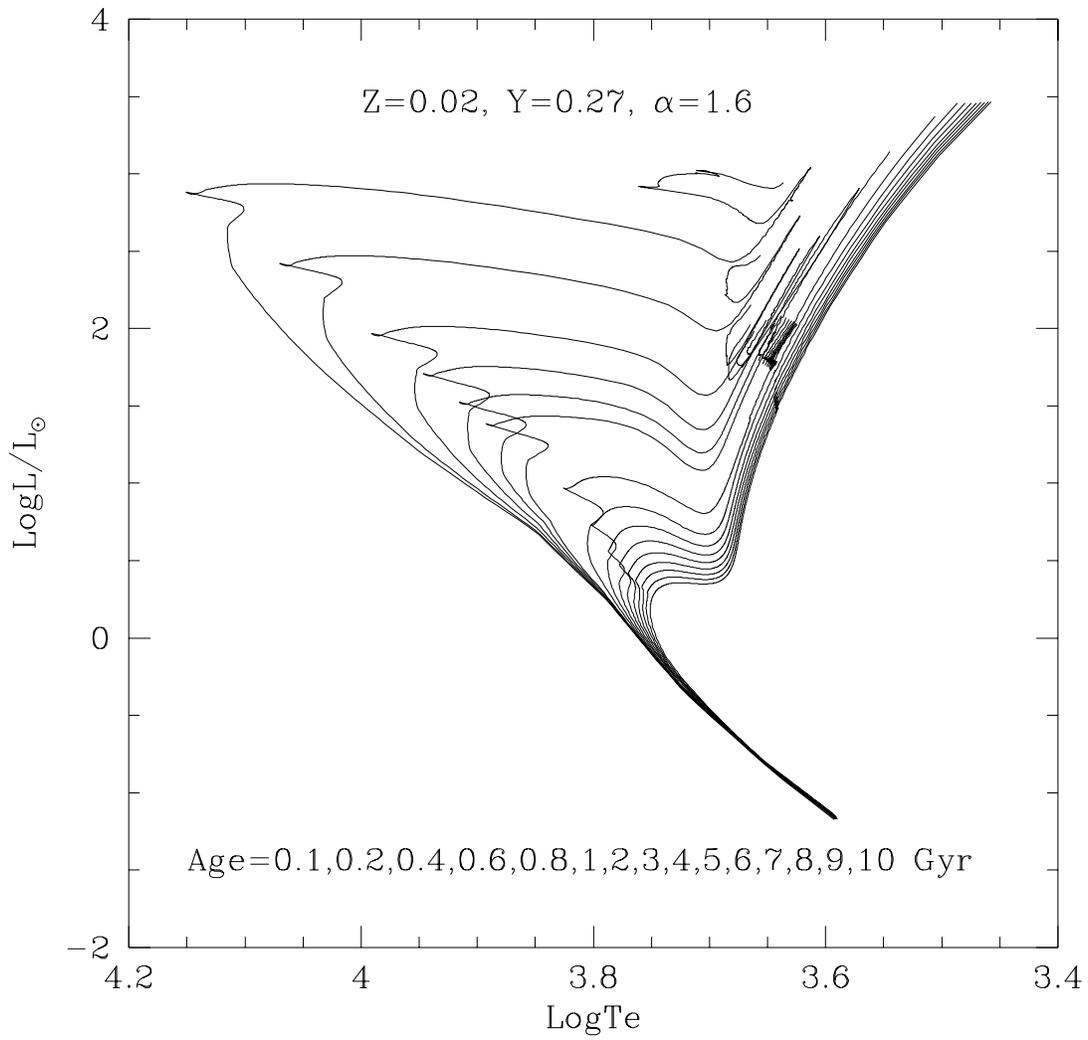
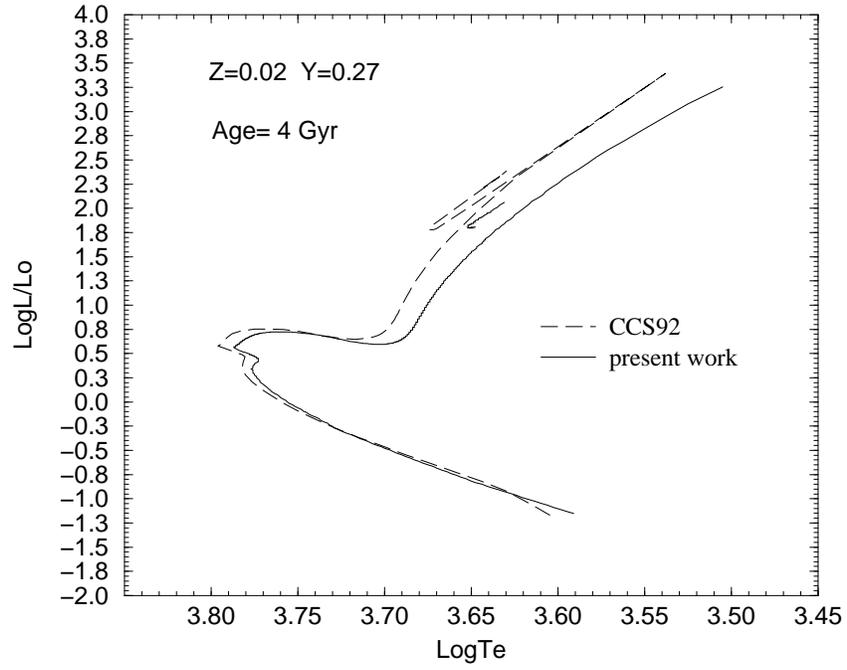
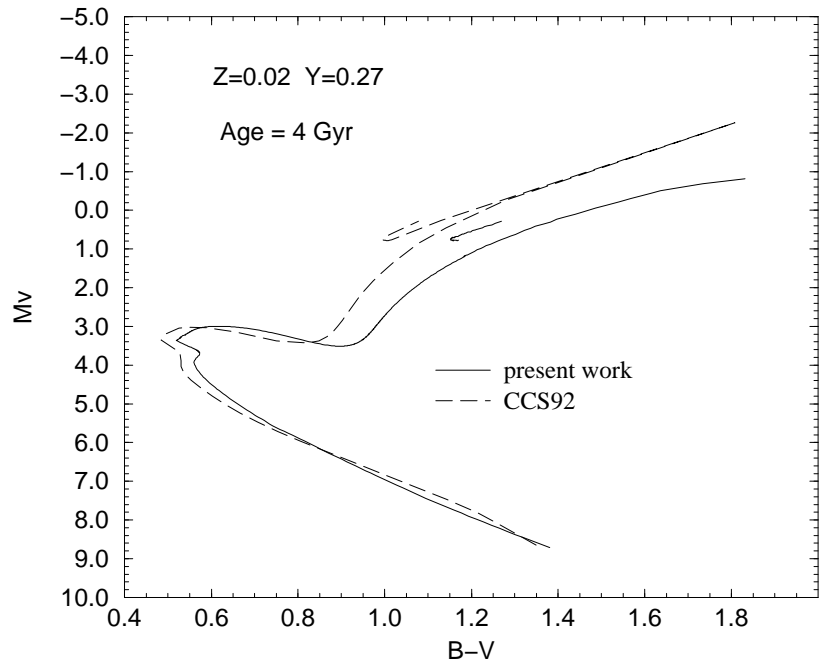
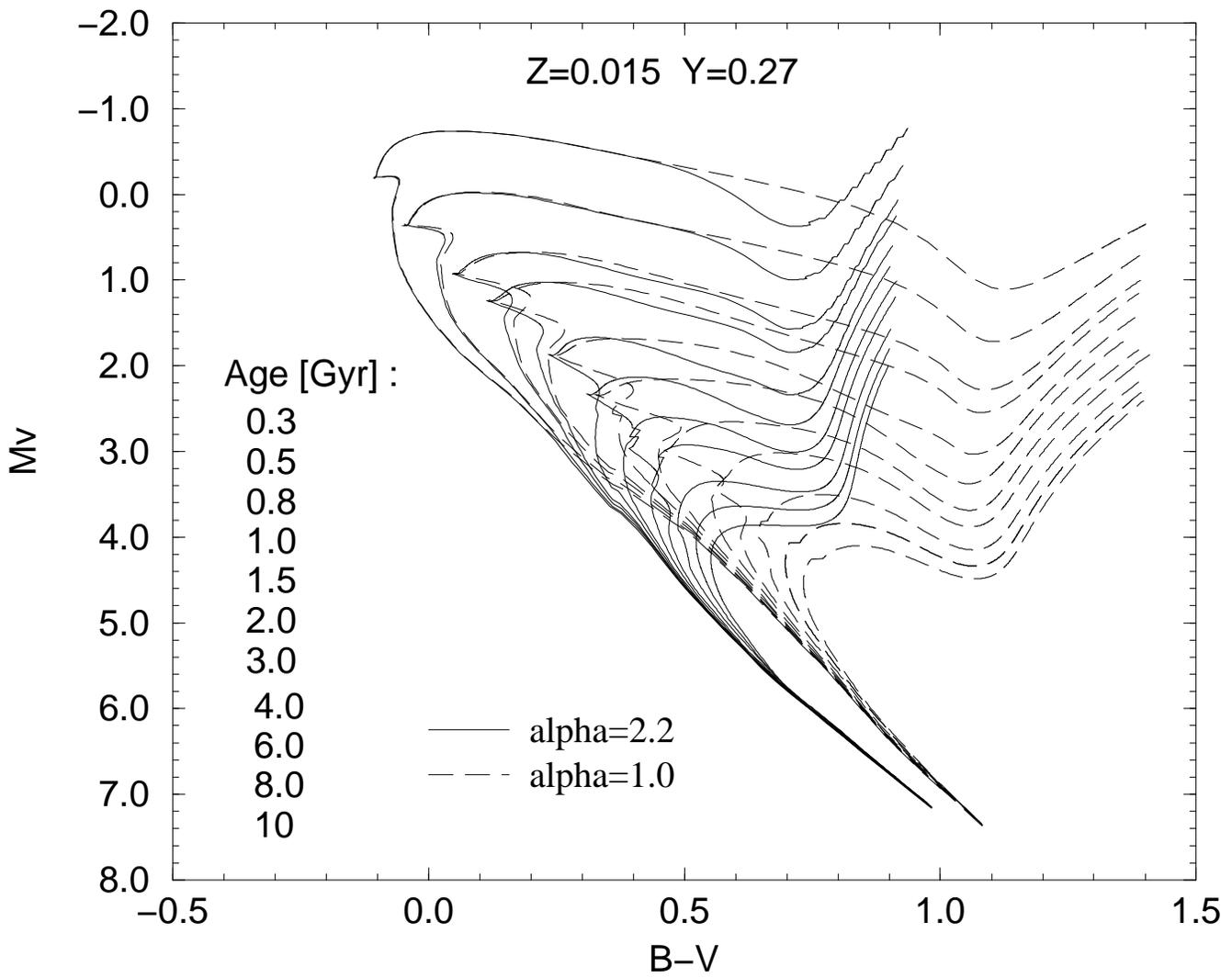
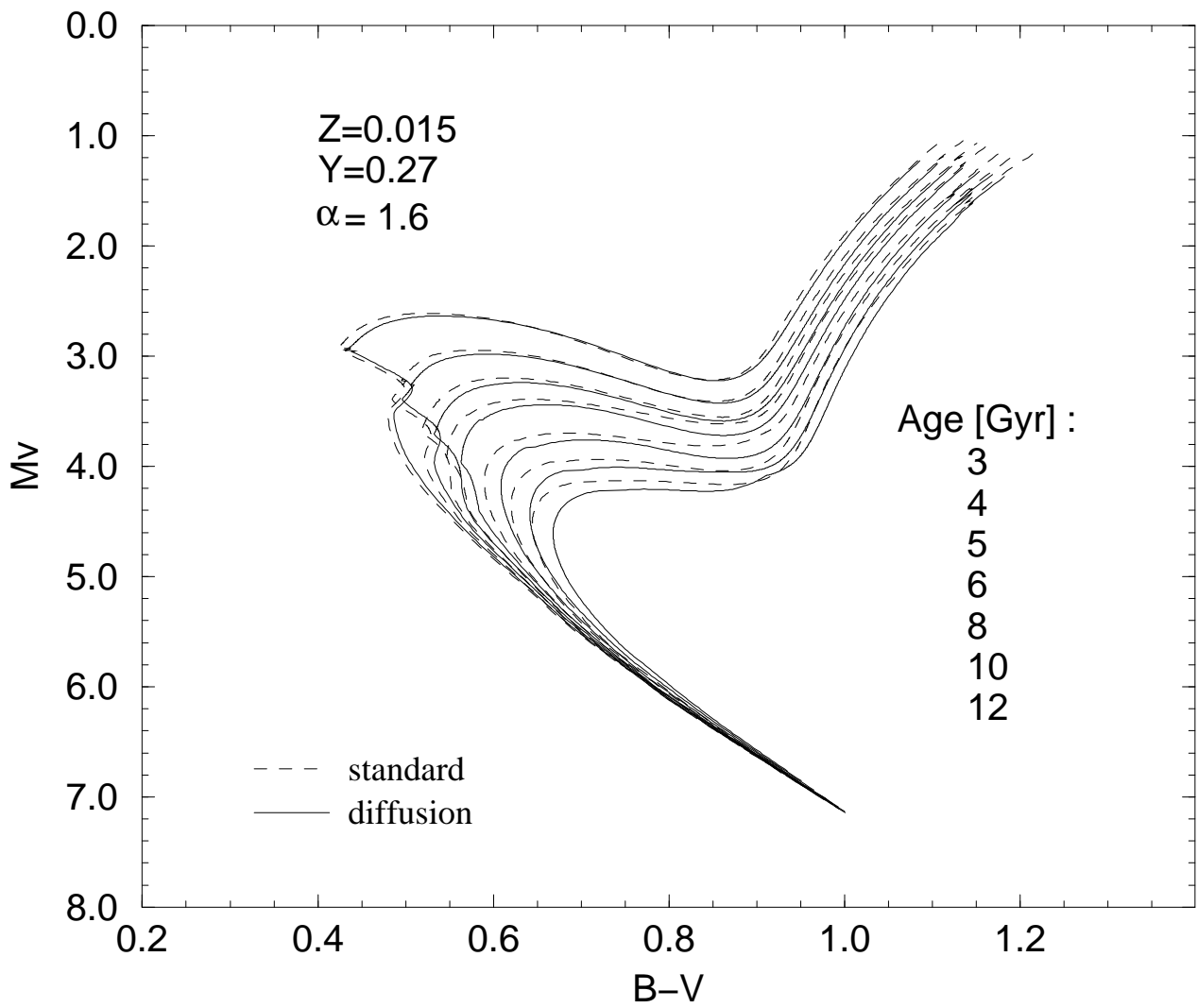
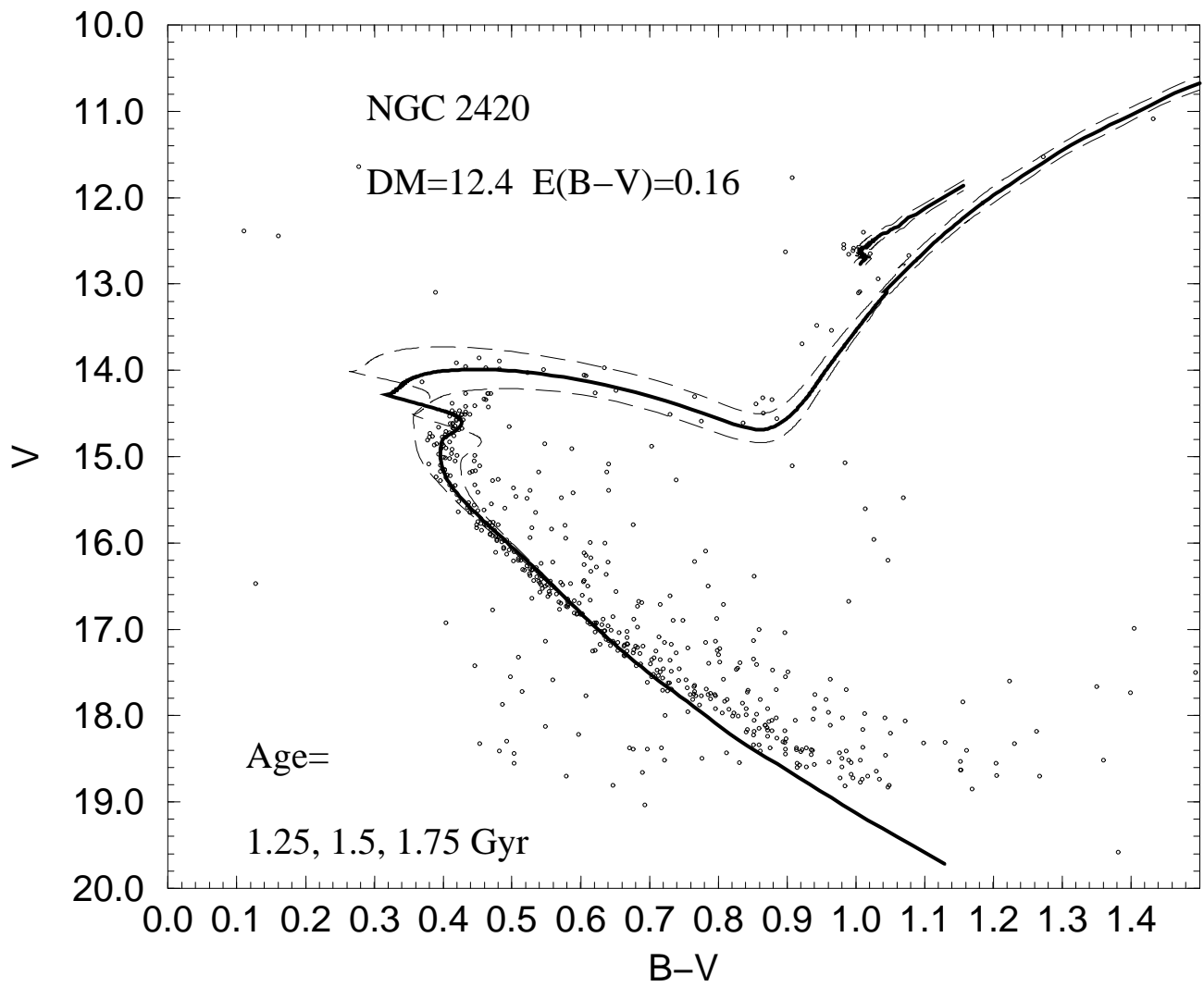
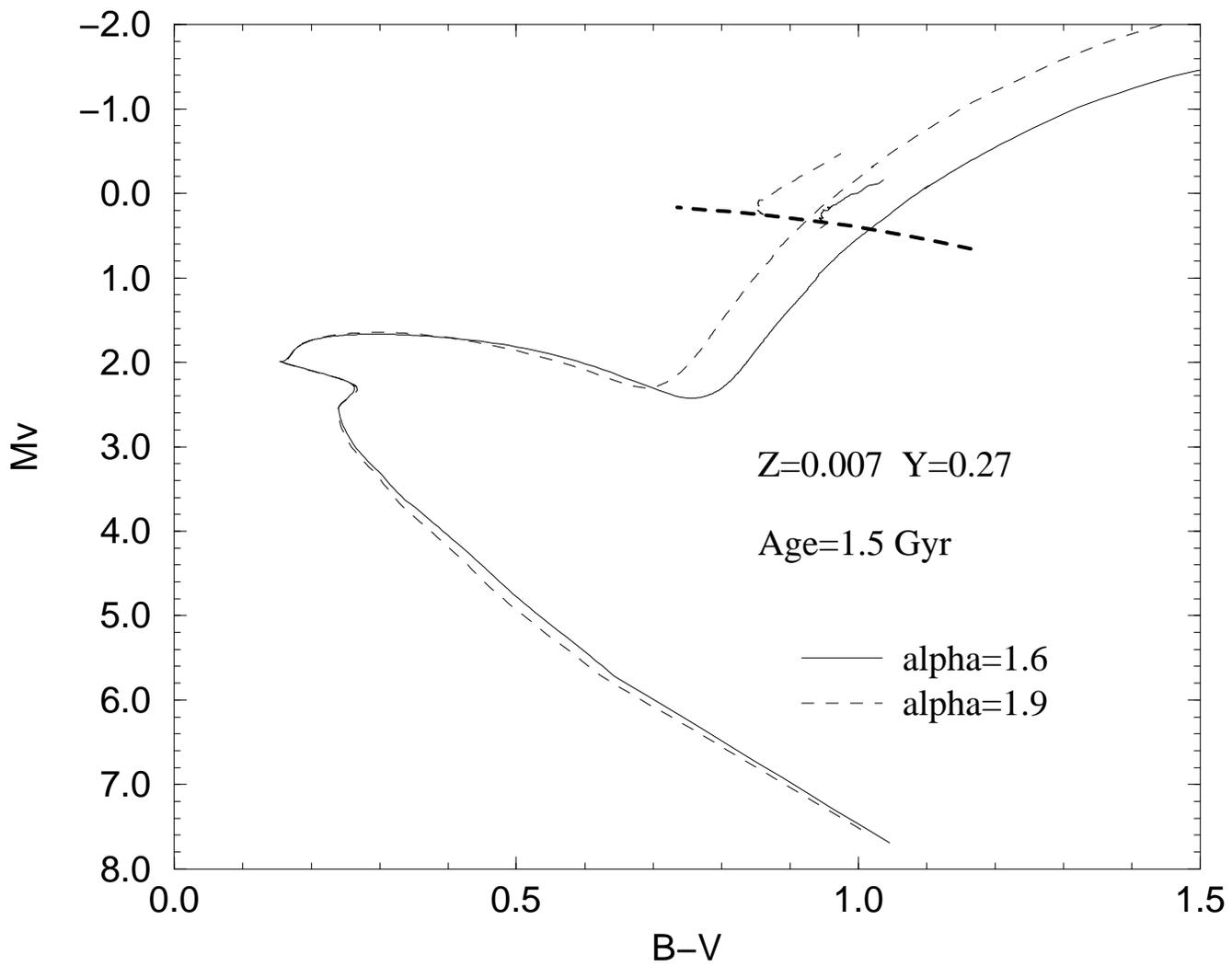
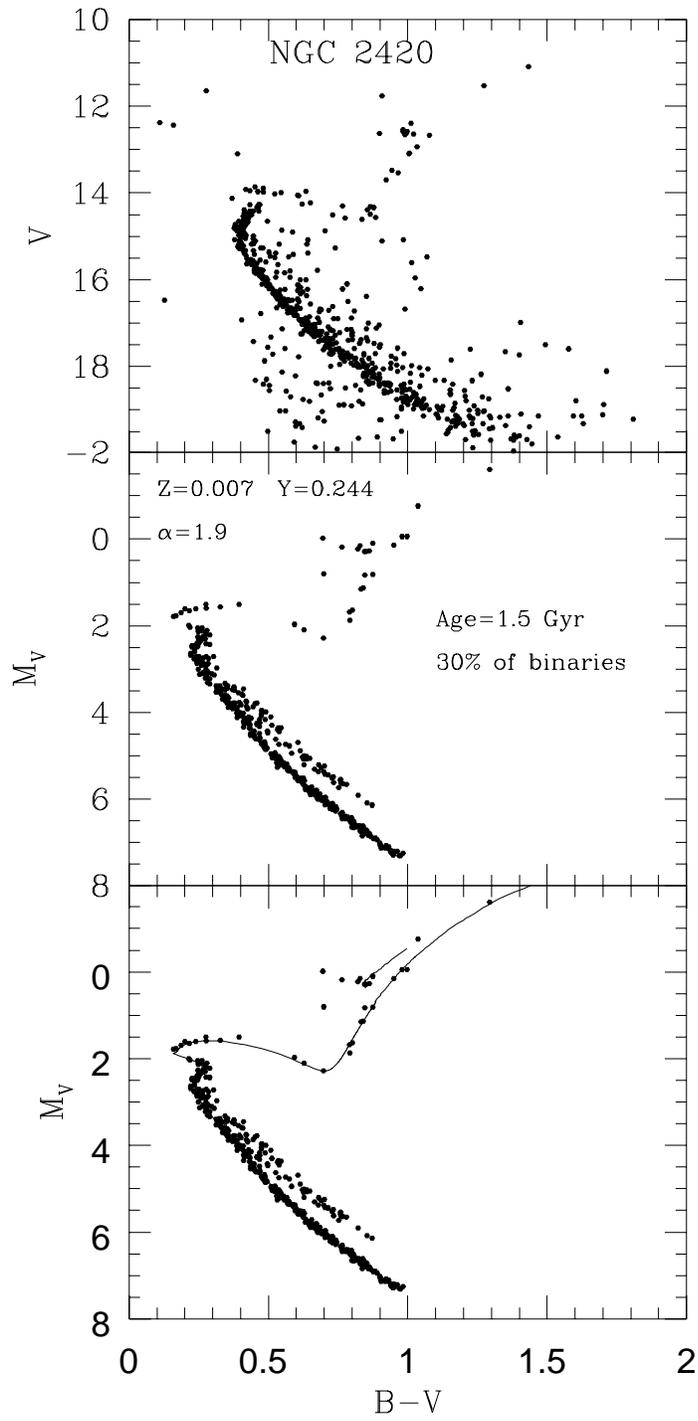
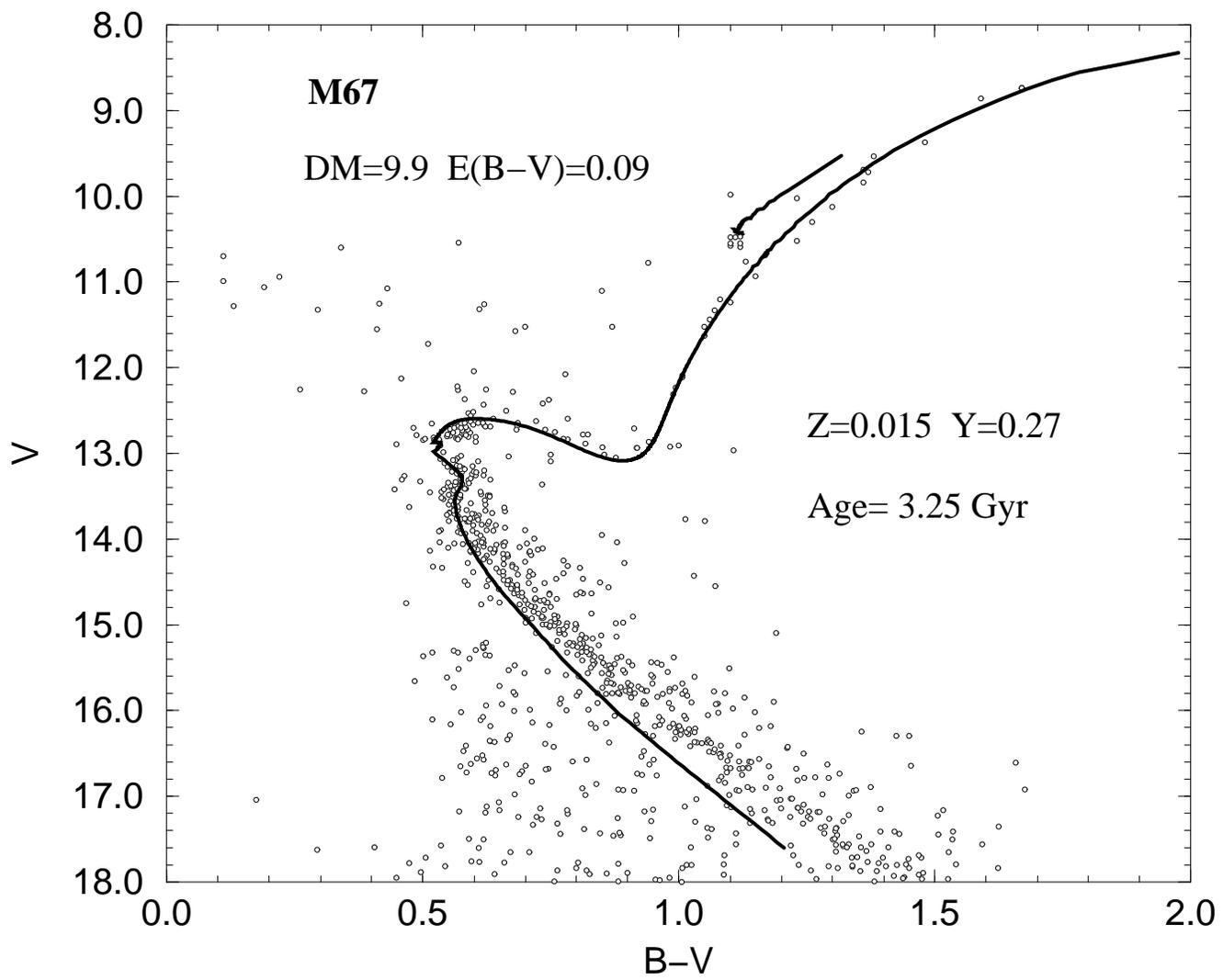
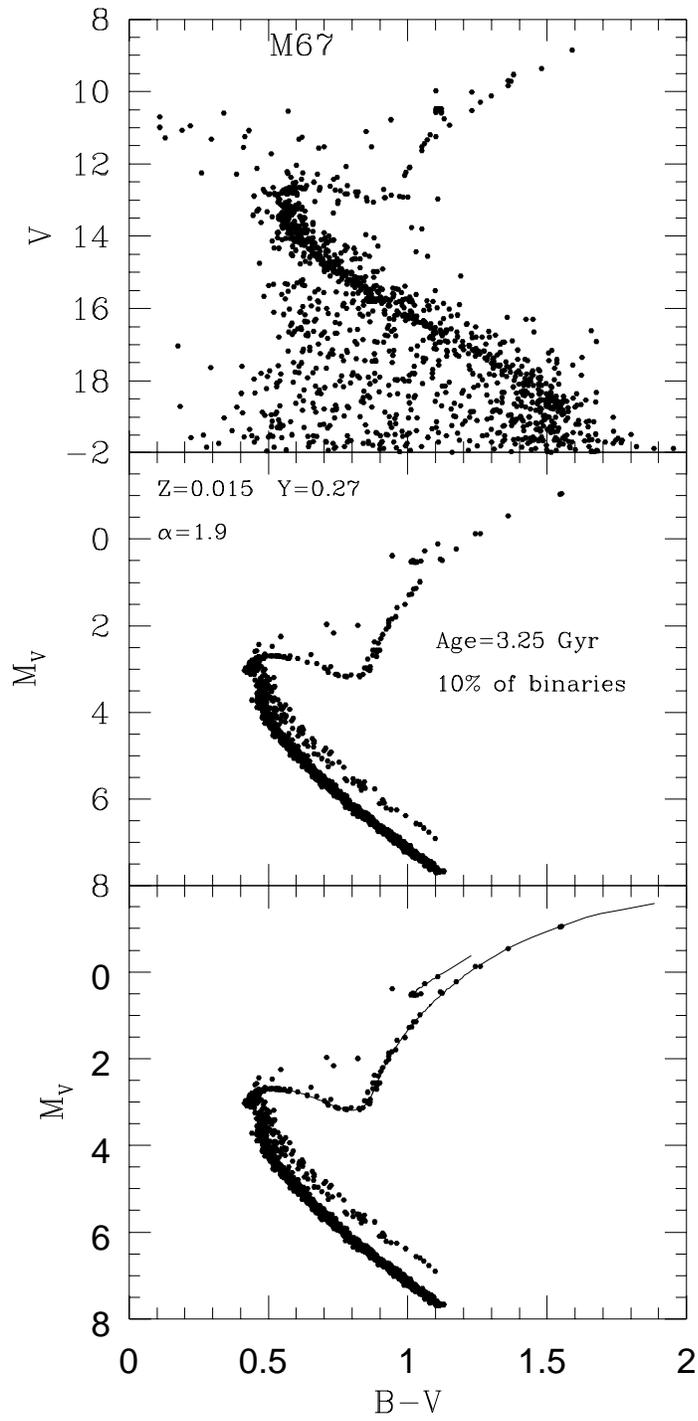
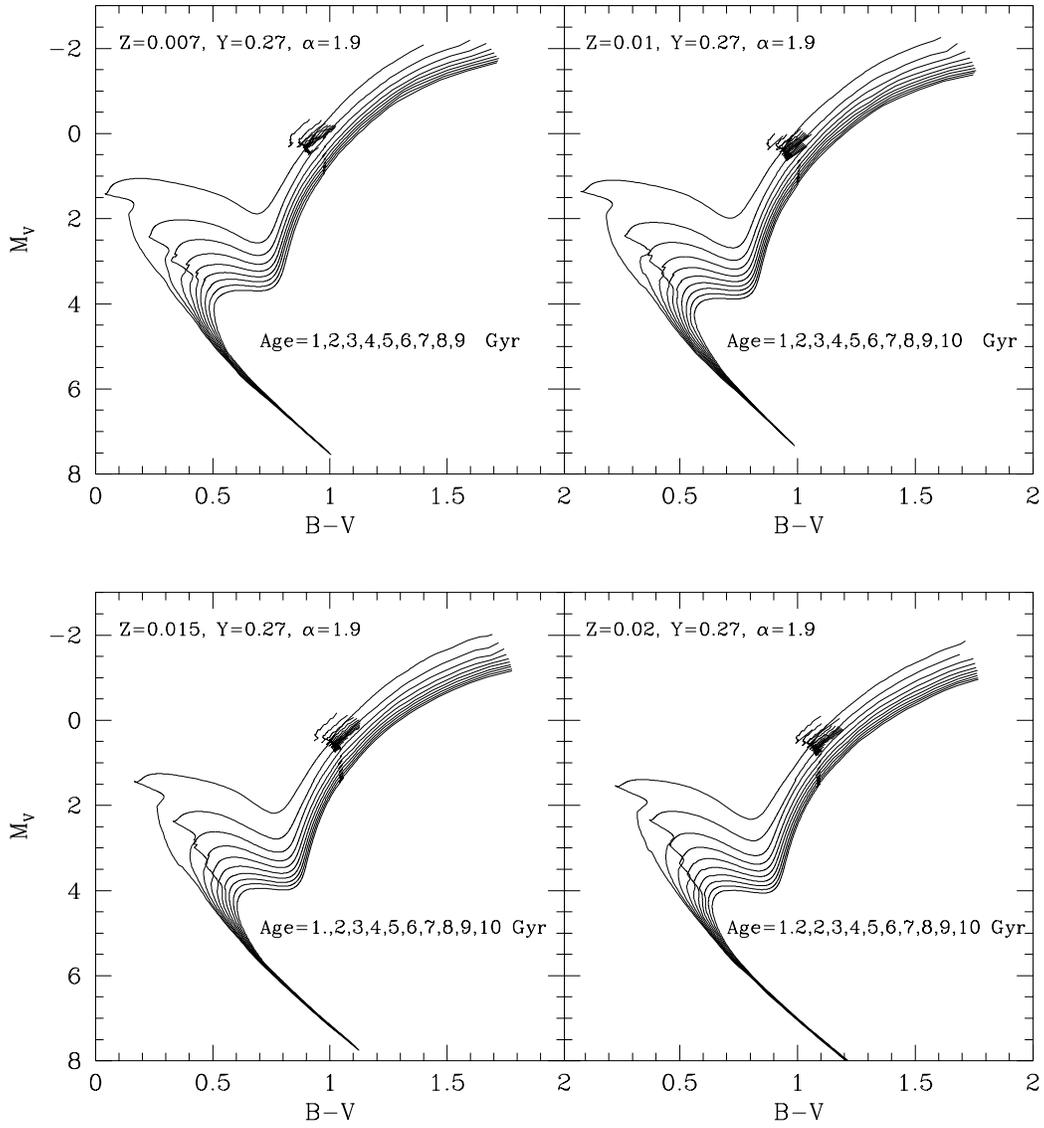
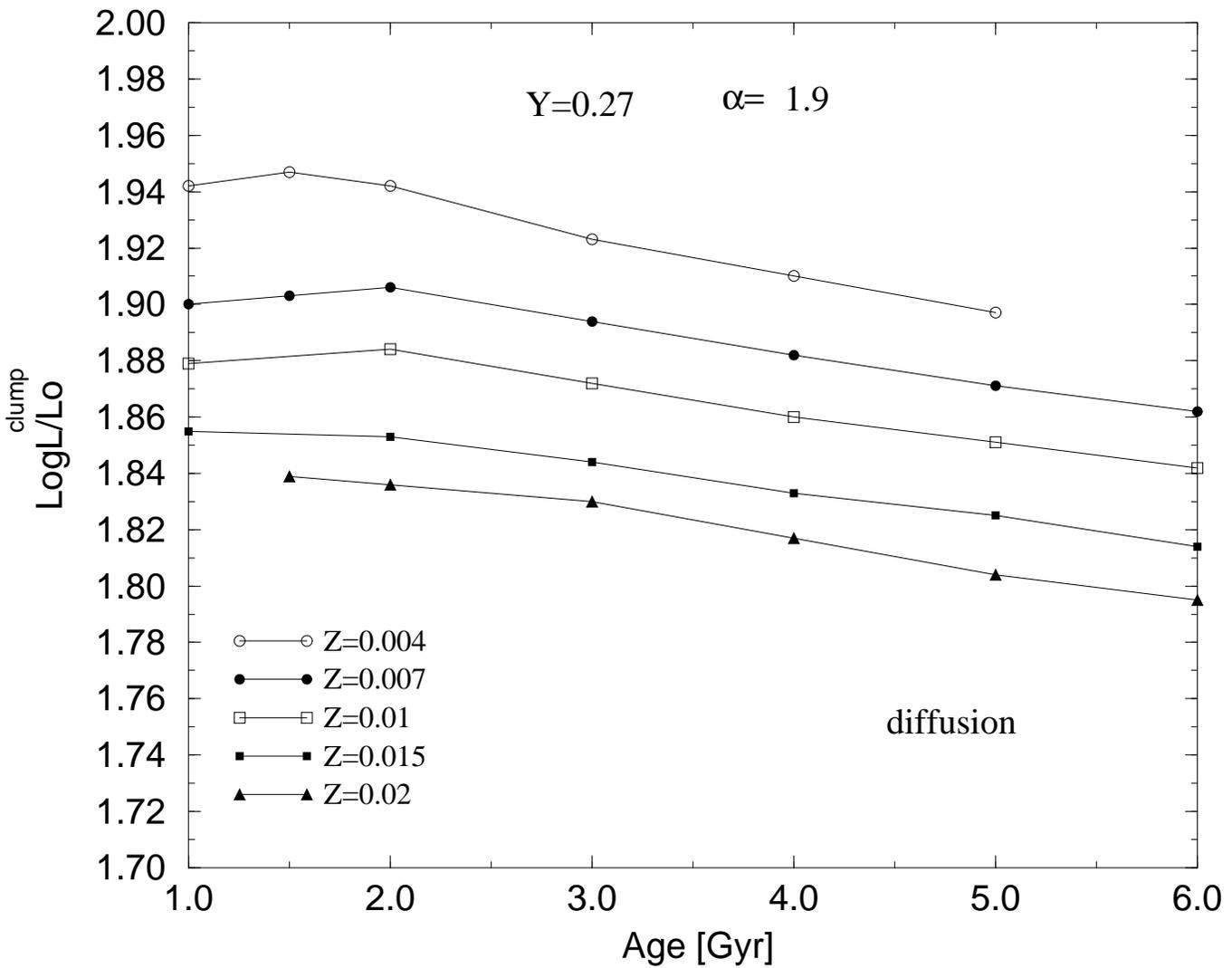
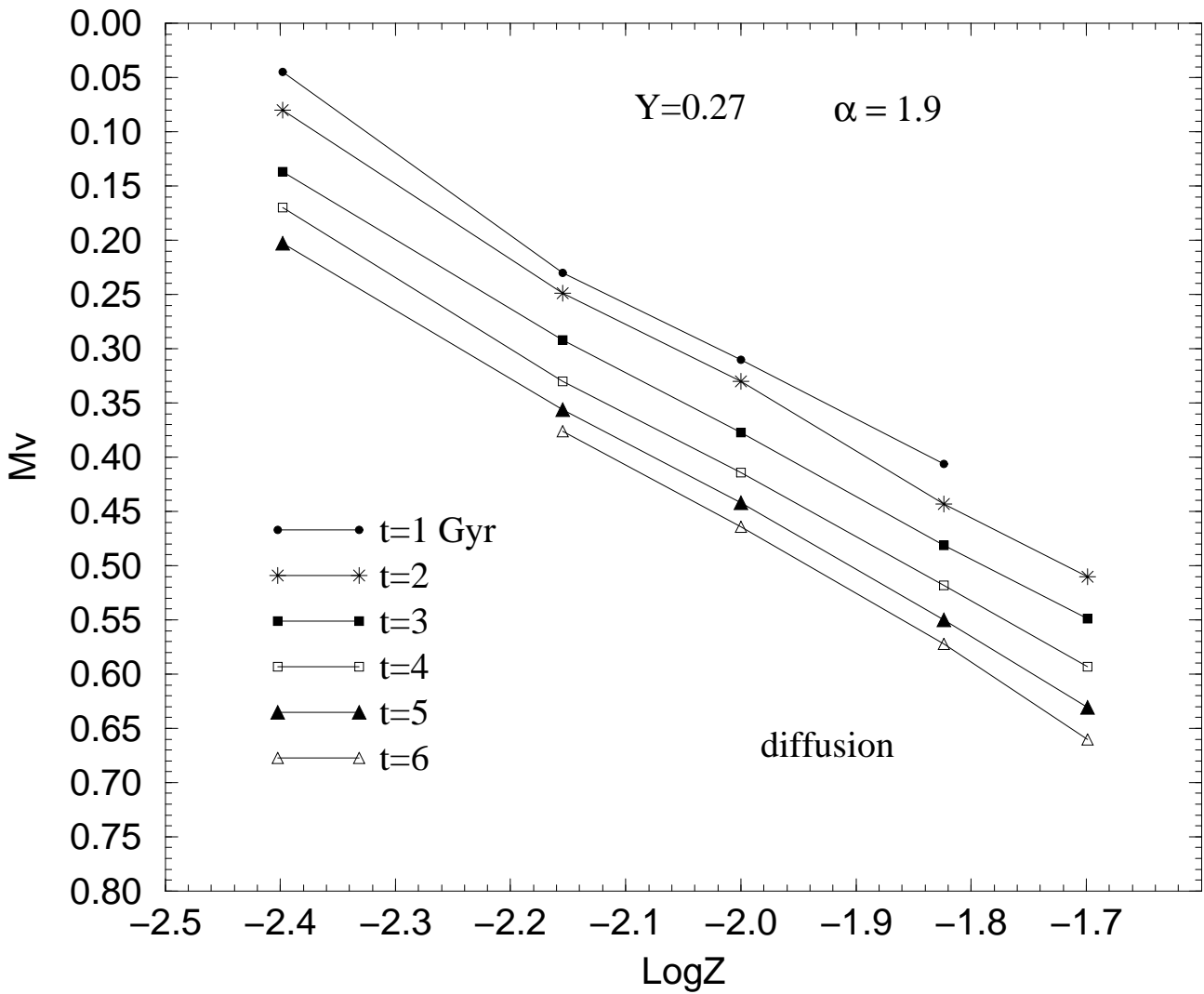
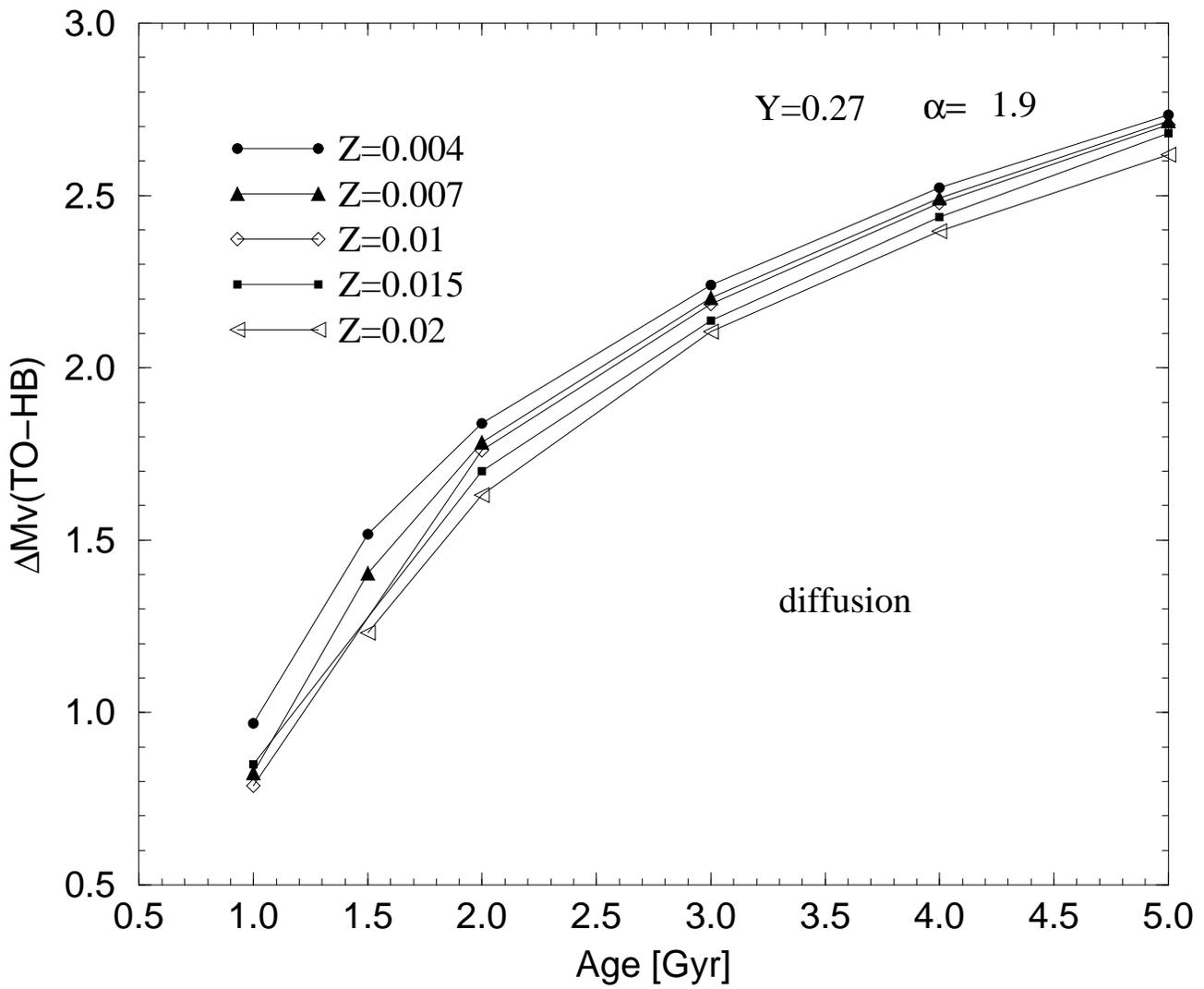
\end{document}